\documentclass[notitlepage,amssymb,amsmath,floatfix,nofootinbib,superscriptaddress]{revtex4-2}
\usepackage{bm}
\usepackage{booktabs}
\usepackage{color}
\usepackage{epsfig}
\usepackage[colorlinks,linkcolor=blue,urlcolor=blue,anchorcolor=black,citecolor=blue]{hyperref}
\usepackage{slashed}
\usepackage{ulem}
\usepackage{xcolor}
\allowdisplaybreaks[4]

\begin{document}

\title{Perturbative QCD Analysis for the Electric Dipole Moment of $\Xi$ and $\Sigma$ Hyperons from CP-Violating Dipole Interactions}

\author{Xi-Yu Mu}   
\affiliation{School of Science, Shandong Jianzhu University, Jinan, Shandong 250101, China}

\author{Kai-Bao Chen}
\email{chenkaibao19@sdjzu.edu.cn}
\affiliation{School of Science, Shandong Jianzhu University, Jinan, Shandong 250101, China}

\author{Jian-Ping Ma}
\email{majp@itp.ac.cn}
\affiliation{School of Physics, Henan Normal University, Xinxiang, Henan 453007, China}
\affiliation{Institute of Theoretical Physics, P.O. Box 2735, Chinese Academy of Sciences, Beijing 100190, China}
\affiliation{School of Physics and Center for High-Energy Physics, Peking University, Beijing 100871, China}

\author{Xuan-Bo Tong}
\email{xutong@jyu.fi}
\affiliation{Department of Physics, University of Jyväskylä, P.O. Box 35, 40014 University of Jyväskylä, Finland}
\affiliation{Helsinki Institute of Physics, P.O. Box 64, 00014 University of Helsinki, Finland}

\begin{abstract}
Hyperon electric dipole moments (EDMs) probe CP-violating interactions beyond the Standard Model. Motivated by the recent BESIII measurement~[\href{https://www.science.org/doi/10.1126/science.adx7334}{Ablikim et al., Science 393, 1027}] and the perturbative QCD analysis of the $\Lambda$ EDM form factor $d_\Lambda$~[\href{https://journals.aps.org/prl/abstract/10.1103/1x3y-qbcx}{Chen et al., Phys. Rev. Lett. 136, 051902}], we extend the perturbative QCD calculation of the EDM form factor to the $\Xi$ and $\Sigma$ hyperons. We derive factorization formulas from quark EDM and chromoelectric dipole moment contributions in terms of leading-twist light-cone distribution amplitudes (LCDAs). Using lattice-QCD results for the LCDAs and the strange-quark dipole moment ranges inferred from the BESIII $d_\Lambda$ bound and neutron EDM constraints, we obtain the conditional estimates $-4.8 \times 10^{-18} < d_{\Xi^-} < 1.8\times10^{-18}\,e\,\mathrm{cm}$ and $-6.1\times10^{-19} < d_{\Sigma^+} < 2.4\times10^{-19}\,e\,\mathrm{cm}$ at the scale of ${J/\psi}$ mass. Dedicated searches at BESIII and the future Super Tau-Charm Factory could further constrain these quark-level sources of CP violation.
\end{abstract}
\maketitle

\section{Introduction}
CP violation is essential for generating the matter-antimatter asymmetry of the Universe~\cite{Sakharov:1967dj}.
However, the CP violation encoded in the Cabibbo-Kobayashi-Maskawa (CKM) quark-mixing matrix of the Standard Model (SM) is far too small to account for the observed asymmetry, motivating searches for additional sources of CP violation. 
A nonzero electric dipole moment (EDM) of a particle signals CP violation.
The strong suppression of EDMs induced by the CKM phase makes EDM measurements sensitive probes of new CP-violating interactions beyond the SM~\cite{Pospelov:2005pr,Engel:2013lsa,Chupp:2017rkp,Alarcon:2022ero}.

The EDMs of the neutron and the electron have been extensively studied and the measured EDM bounds impose stringent constraints~\cite{Abel:2020pzs,Roussy:2022cmp}.
In contrast, experimental measurements of EDMs of other particles, including hyperons, are scarce.
However, these measurements are crucial for unraveling the underlying interactions at the quark or gluon level as well as for flavor separation.
Strange baryons offer such complementary information. An early upper bound on the static EDM of the $\Lambda$ hyperon was obtained at Fermilab in the 1980s~\cite{Pondrom:1981gu}.

The weak decays of hyperons analyze their spins, making correlated baryon-antibaryon pairs particularly useful for CP tests.
Proposals exploiting this feature in charmonium decays~\cite{He:1992ng,He:2022jjc} have developed into detailed studies of production amplitudes, angular observables, and experimental sensitivities~\cite{He:2022jjc,Fu:2023ose,Du:2024jfc,Guo:2025bfn}.
Recently, the BESIII Collaboration extracted the $\Lambda$ EDM form factor from entangled $\Lambda\bar{\Lambda}$ pairs in $J/\psi$ decays~\cite{BESIII:2025vgs}.
The resulting bound is numerically nearly three orders of magnitude smaller than the earlier static EDM limit~\cite{Pondrom:1981gu}, which is a significant improvement.
For a proper theoretical interpretation, however, it is important to note that the two measurements probe different kinematic regimes. BESIII constrains the timelike EDM form factor $d_\Lambda(Q)$ at $Q=M_{J/\psi}$, the invariant mass of the $\Lambda\bar{\Lambda}$ pair, whereas the earlier experiment constrained the static EDM $d_\Lambda(0)$.


The invariant mass $Q=M_{J/\psi}$ of the $\Lambda\bar{\Lambda}$ pair provides a hard scale relative to $\Lambda_{\mathrm{QCD}}$, motivating a perturbative QCD treatment of the EDM form factor.
In Ref.~\cite{Chen:2025rab}, the contributions of quark EDMs and chromoelectric dipole moments (CEDMs) to $d_\Lambda(Q)$ were calculated within the leading-twist collinear factorization framework for hard exclusive processes~\cite{Avdeenko:1981twg,Lepage:1979za,Brodsky:1981kj,Efremov:1979qk,Chernyak:1977as,Chernyak:1980dj,Lepage:1980fj,Chernyak:1983ej,Chernyak:1984bm}. 
This framework has also been widely applied to hadronic electromagnetic and gravitational form factors~\cite{Belitsky:2002kj,Sun:2021gmi,Sun:2021pyw,Tong:2021ctu,Tong:2022zax,Chen:2024fhj,Huang:2024ugd,Chen:2023byr,Ji:2024iak,Ding:2024lfj,Chen:2024oem,Coriano:2024qbr,Wang:2025irh}. 
The dipole interactions provide the necessary quark helicity flip, allowing nonzero contributions at the leading twist. 
The form factor can therefore be expressed as convolutions of perturbatively calculable hard coefficients with nonperturbative light-cone distribution amplitudes (LCDAs). 
Numerical predictions reveal particular sensitivity to the strange-quark CEDM, providing information complementary to neutron EDM constraints.

On the experimental side, BESIII has collected approximately $10^{10}$ $J/\psi$ events~\cite{BESIII:2021cxx}, providing large samples of spin-correlated hyperon--antihyperon pairs.
The parity-violating weak decays of hyperons act as spin analyzers, allowing their polarization and spin correlations to be reconstructed from the joint angular distributions of the decay products~\cite{Zheng:2025tnz}.
This strategy has enabled polarization measurements and CP tests in the $\Lambda$, $\Sigma$, and $\Xi$ channels~\cite{BESIII:2018cnd,BESIII:2022qax,BESIII:2020fqg,BESIII:2021ypr,BESIII:2023drj}, with recent analyses further improving the precision of the $\Sigma^+$ decay asymmetry parameters and the $\Xi^-$ decay parameters and weak phases~\cite{BESIII:2025jxt,BESIII:2026hgj}.
Together with the recent extraction of the $\Lambda$ EDM form factor~\cite{BESIII:2025vgs}, these advances provide an experimental basis for extending dedicated searches to the timelike EDM form factors of $\Sigma$ and $\Xi$ hyperons.
The possibility of probing these EDMs through angular analyses of $J/\psi$ decays has also been explored~\cite{Zhang:2010zzo}.

The theoretical motivation of our study follows from the complementary flavor sensitivities of different hyperons.
The distinct quark content and spin-flavor structures of $\Lambda$, $\Xi$, and $\Sigma$ lead to different combinations of quark dipole contributions and therefore provide additional information on the flavor structure of CP-violating interactions.
A common factorization framework also permits a consistent treatment among different hyperons.
Comparisons across hyperon channels can also test the applicability of this factorization framework.

Motivated by these experimental and theoretical aspects, we extend our previous perturbative-QCD analysis of the $\Lambda$ EDM form factor~\cite{Chen:2025rab} to the $\Xi$ and $\Sigma$ hyperons, providing quantitative estimates for interpreting future experimental measurements.

The rest of this paper is organized as follows.
In Sec.~\ref{sec:framework}, we set up the framework and introduce the hyperon LCDAs.
In Sec.~\ref{sec:pQCDcalc}, we present a perturbative QCD calculation of the hyperon EDM form factor and derive the factorized analytical results.
Sec.~\ref{sec:numerical} presents a numerical analysis.
A summary is given in Sec.~\ref{sec:summary}.

\section{Framework and hyperon LCDAs}
\label{sec:framework}
\subsection{CP-violating dipole interactions and the EDM form factor}
We consider the EDMs of hyperons generated by quark dipole interactions, which are introduced through a CP-violating effective Lagrangian as
\begin{align}
\mathcal L_{\mathrm{CP}} = -\frac{i}{2}\sum_{q=u,d,s}\bigl(
d_q\bar q \gamma_5\sigma^{\mu\nu}q F_{\mu\nu} + \tilde d_q\bar q\gamma_5\sigma^{\mu\nu} q G_{\mu\nu} \bigr).
\label{eq:lagrangian}
\end{align}
Here, $d_q$ and $\tilde d_q$ denote the quark EDM and CEDM,
respectively.
$q$ denotes the quark field of flavor $u$, $d$, and $s$.
$F_{\mu\nu}$ is the photon field strength tensor, and $G_{\mu\nu}\equiv G_{\mu\nu}^c T^c$ is the gluon field strength tensor. We use $\sigma^{\mu\nu}=i[\gamma^\mu,\gamma^\nu]/2$.

CP violation is indicated by the presence of the EDM form factor defined through the matrix element
\begin{align}
&{\cal M}^\mu = \langle0|J^\mu|H(P,\lambda)\bar H(\bar P,\bar\lambda)\rangle_{\rm cpv} = d_H(Q)\bar V_{\bar H}(\bar P,\bar\lambda)\gamma_5\sigma^{\mu\nu}q_\nu U_H(P,\lambda),
\label{eq:vertex}
\end{align}
where $H(P,\lambda)$ denotes a hyperon $H$ with momentum $P$ and helicity $\lambda$, and $\bar H(\bar P,\bar \lambda)$ denotes the corresponding anti-hyperon. The invariant mass of the $H\bar H$ system is $Q=\sqrt{q^2}$ with $q=P+\bar P$.
$J^\mu$ is the electromagnetic current. $U_H$ and $\bar V_{\bar H}$ are the spinor of the hyperon and the anti-hyperon, respectively.
The subscript ``$\rm cpv$" stands for taking the CP-violating part of the matrix element.
The CP-conserving contributions of the matrix element are not shown.
The coefficient $d_H(Q)$ is called the EDM form factor.
A nonzero $d_H(Q)$ indicates the existence of CP violation.
It should be clear that $d_H(Q)$ differs from the static hyperon EDM at $Q=0$. It depends on the invariant mass of the hyperon system.

The static neutron EDM has been extensively studied within low-energy QCD effective field theories or lattice QCD~\cite{Chupp:2017rkp,Alarcon:2022ero}, and the measured EDM bounds impose a stringent constraint~\cite{Abel:2020pzs}.
There are quark model predictions for hyperon EDMs from quark EDMs and CEDMs in the non-relativistic QCD limit~\cite{Du:2024jfc}.
However, measurements for hyperon EDMs are scarce since hyperons are unstable particles.
The existing measurements almost exclusively rely on particle decay processes at energy scales much larger than the QCD nonperturbative scale.
The decay of $J/\psi$ meson into hyperon-antihyperon pairs serves as the primary channel, where the EDM form factor corresponds to the value evaluated at $Q=M_{J/\psi}$.
At this scale, the calculation of $d_H(Q)$ can be done within the framework of collinear factorization for hard exclusive processes.
This calculation has been carried out for the $\Lambda$ hyperon~\cite{Chen:2025rab} which is an isospin-0 baryon.
We extend the calculation to hyperons with isospin-1/2 and isospin-1, i.e., $\Xi$ and $\Sigma$ hyperons.
For particles within a given isospin multiplet, the calculations are quite similar. Therefore, we will explicitly evaluate only one representative from each multiplet in this paper.
Specifically, we take $\Xi^-$ and $\Sigma^+$ as examples for the $\Xi$ doublet and the $\Sigma$ triplet, respectively.

\subsection{LCDAs for $\Xi^-$}
We first consider the case of $\Xi^-$, whose valence quark composition is $(ssd)$. We proceed by introducing the LCDAs of $\Xi^-$.
Throughout this paper, we use the light-cone coordinate system, in which a vector $a^\mu$ is expressed as $a^\mu = (a^+,a^-, \boldsymbol{a}_\perp) = \bigl((a^0+a^3)/{\sqrt{2}}, (a^0-a^3)/{\sqrt{2}}, a^1, a^2 \bigr)$.
We also define two light-cone vectors $n^\mu = (0,1,\boldsymbol{0}_\perp)$ and $l^\mu = (1,0,\boldsymbol{0}_\perp)$.
For $\Xi^-(P,\lambda) \bar \Xi^+(\bar P,\bar \lambda) \to \gamma^*(q)$, in the center of mass frame of $\Xi^-$ and $\bar \Xi^+$ and neglecting the hadron mass, the momenta are given by $P^\mu = (Q/\sqrt{2},0,\boldsymbol{0}_\perp)$, and $\bar P^\mu = (0,Q/\sqrt{2},\boldsymbol{0}_\perp)$.
The leading-twist LCDAs for $\Xi^-$ in the momentum space are defined as~\cite{Chernyak:1987nu,Braun:1999te}
\begin{align}
& 4\varepsilon^{abc} \int \frac{d\xi_1 d\xi_2 d\xi_3}{(2\pi)^3} e^{iP^+(x_1\xi_1 + x_2\xi_2 + x_3\xi_3)} \langle 0| s_\alpha^a\left(\xi_1 n\right) s_\beta^b\left(\xi_2 n\right) d_\gamma^c\left(\xi_3 n\right)\left| \Xi^-(P)\right\rangle \nonumber \\
&= (\gamma\cdot P C)_{\alpha \beta}\left(\gamma_5 U_{\Xi^-} \right)_\gamma {\cal V}_{{\Xi^-}}\left([x]\right) + \left(\gamma\cdot P \gamma_5 C\right)_{\alpha \beta} \left(U_{\Xi^-}\right)_\gamma {\cal A}_{{\Xi^-}}\left([x]\right) + \left( P_{\nu} i\sigma^{\mu \nu} C\right)_{\alpha \beta}\left(\gamma_\mu \gamma_5 U_{\Xi^-}\right)_\gamma {\cal T}_{{\Xi^-}}\left([x]\right) + \cdots, \label{eq:Xi-LCDA}
\end{align}
where $C$ is the charge-conjugation matrix.
The $\cdots$ stand for terms of higher-twist.
The coefficient functions ${\cal V}_{\Xi^-}$, ${\cal A}_{\Xi^-}$, and ${\cal T}_{\Xi^-}$ are called as the LCDAs.
They depend on the momentum fractions, $x_1$, $x_2$, and $x_3$, of the hadron carried by the three valence quarks.
The momentum fraction dependence is denoted collectively as $[x]$.
Here, we have already absorbed the normalization constants into the definition of the LCDAs, so that the LCDAs have mass dimension two.
Under the exchange of the momentum fractions carried by the two $s$ quarks, i.e., $x_1\leftrightarrow x_2$, these LCDAs satisfy~\cite{Chernyak:1987nu}
\begin{align}
& {\cal V}_{\Xi^-}\left(x_1, x_2, x_3\right) = {\cal V}_{\Xi^-}\left(x_2, x_1, x_3\right), \label{eq:symV}\\
& {\cal A}_{\Xi^-}\left(x_1, x_2, x_3\right)=-{\cal A}_{\Xi^-}\left(x_2, x_1, x_3\right), \label{eq:symA} \\
& {\cal T}_{\Xi^-}\left(x_1, x_2, x_3\right)={\cal T}_{\Xi^-}\left(x_2, x_1, x_3\right). \label{eq:symT}
\end{align}

The physical meaning of these LCDAs can be seen from the $\Xi^-$ state with three valence quarks in helicity basis~\cite{Bali:2015ykx},
\begin{align}
\left|\Xi^{-\uparrow}\right\rangle= & \int_0^1 \frac{[d^3x]}{8 \sqrt{6x_1x_2x_3}}\Bigl\{ \Bigl[ {\cal V}_{\Xi^-}([x]) +  {\cal A}_{\Xi^-}([x])\Bigr] \left|s^{\downarrow} s^{\uparrow} d^{\uparrow} \right\rangle 
+ \Bigl[ {\cal V}_{\Xi^-}([x]) -  {\cal A}_{\Xi^-}([x])\Bigr] \left|s^{\uparrow} s^{\downarrow} d^{\uparrow} \right\rangle 
- 2 {\cal T}_{\Xi^-}([x])\left|s^{\uparrow} s^{\uparrow} d^{\downarrow} \right\rangle\Bigr\},\label{eq:XiWF}
\end{align}
where $[d^3x] \equiv dx_1 dx_2 dx_3 \delta(1-x_1-x_2-x_3)$.
The up- or down-arrow in the superscript denotes for the $+$ or $-$ helicity state, respectively.
Through charge-conjugation transformation, one can get the definition of the $\bar\Xi^+$ LCDAs as
\begin{align}
& 4\varepsilon^{abc} \int \frac{d\xi_1 d\xi_2 d\xi_3}{(2\pi)^3} e^{i\bar P^-(x_1\xi_1 + x_2\xi_2 + x_3\xi_3)} \langle 0| \bar s_\alpha^a\left(\xi_1 l\right) \bar s_\beta^b\left(\xi_2 l\right) \bar d_\gamma^c\left(\xi_3 l\right)\left|\bar \Xi^{+}(\bar P)\right\rangle \nonumber \\
&= (C \gamma\cdot \bar P)_{\alpha \beta}\left( \bar V_{\bar \Xi^+} \gamma_5 \right)_\gamma {\cal V}_{\bar \Xi^+}\left([x]\right) + \left(C \gamma\cdot \bar P \gamma_5 \right)_{\alpha \beta} \left(\bar V_{\bar \Xi^+}\right)_\gamma {\cal A}_{\bar \Xi^+}\left([x]\right) + \left( \bar P_{\nu} C i\sigma^{\mu \nu} \right)_{\alpha \beta}\left( \bar V_{\bar \Xi^+} \gamma_\mu \gamma_5 \right)_\gamma {\cal T}_{\bar \Xi^+}\left([x]\right) + \cdots. \label{eq:XibarLCDA}
\end{align}

The LCDAs are universal inputs relevant to hadronic structure.
They provide the nonperturbative projectors needed for the calculation of the hyperon EDM form factor.
Their factorization-scale dependence is implicit here.

\section{Perturbative calculations}
\label{sec:pQCDcalc}
In this section, we present a perturbative calculation for the hyperon EDM form factor.
We consider contributions from quark EDMs and CEDMs and derive, at tree level and leading twist, a factorized form for the hyperon EDM form factor in terms of the hard coefficients and hyperon LCDAs.

\subsection{Quark EDMs contribution to $d_{\Xi^-}(Q)$}
We first calculate the contributions from quark EDMs.
In the high-energy limit with the hadron mass neglected, because the Dirac structure in Eq.~(\ref{eq:vertex}) flips the hyperon helicity, the amplitude for $\lambda = \bar \lambda$ vanishes due to helicity conservation.
Consequently, only matrix elements with $\lambda \ne \bar\lambda$ need to be considered.
Ordinary QCD and QED interactions conserve massless-quark helicity. 
To generate such a helicity-flipped amplitude, higher-twist effects or orbital angular momentum are usually required.
However, the quark dipole interactions in Eq.~(\ref{eq:lagrangian}) already flip the quark helicities.
As a result, $d_H(Q)$ will receive nonzero contributions at the leading-twist level.
Taking the quark EDMs contribution as an example, based on Eq.~(\ref{eq:XiWF}), one can find that, for the contribution proportional to $d_s$, the nonzero cases will involve the structure of $({\cal V}_{\Xi^-} \pm {\cal A}_{\Xi^-}) \otimes {\cal T}_{\bar \Xi^+}$ or ${\cal T}_{\Xi^-} \otimes ({\cal V}_{\bar \Xi^+} \pm {\cal A}_{\bar \Xi^+})$.
For the contribution proportional to $d_d$, the nonzero cases are $({\cal V}_{\Xi^-} \pm {\cal A}_{\Xi^-}) \otimes ({\cal V}_{\bar \Xi^+} \mp {\cal A}_{\bar \Xi^+})$.
The result can be checked by explicit calculations.
The calculation procedure is similar to the case of $\Lambda$ EDM form factor.
\begin{figure}[!htb]
\centering
\includegraphics[width=0.55\textwidth]{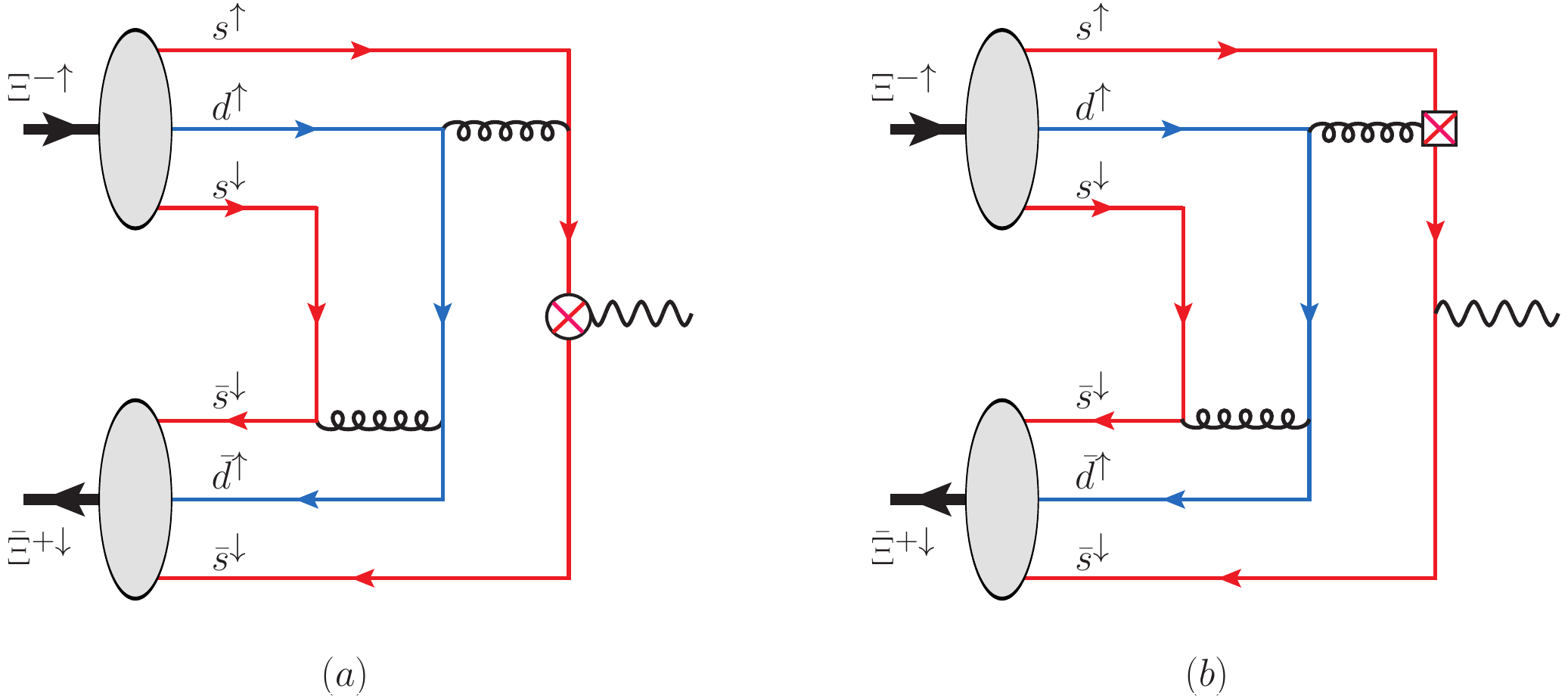}
\caption{Representative diagrams for the $\Xi^-$ EDM form factor with (a) the strange-quark EDM and (b) the strange-quark CEDM. The circled cross denotes an EDM photon vertex, and the boxed cross denotes a CEDM gluon vertex. The top, middle, and bottom quark lines leaving the $\Xi^-$ blob are $s$, $d$, and $s$ quarks with indices 2, 3, and 1, respectively.}
\label{fig:EDM}
\end{figure}

One typical diagram for the contribution from the $s$ quark EDM is shown in Fig.~\ref{fig:EDM}(a).
The amplitude is written as
\begin{align}
-i{\cal M}^\mu_{(a)} &= \int \frac{dk_1^+ dk_2^+ dk_3^+}{(2\pi)^3} \frac{dp_1^- dp_2^- dp_3^-}{(2\pi)^3} d\xi_1 d\xi_2 d\xi_3 d\eta_1 d\eta_2 d\eta_3 e^{ik_1^+\xi_1 + ik_2^+\xi_2 + ik_3^+\xi_3} e^{ip_1^-\eta_1 + ip_2^-\eta_2 + ip_3^-\eta_3} \nonumber\\
&\times (-1) \frac{1}{2!} \langle 0| \bar s_{\beta_1}^{b_1}(\eta_1 l) \bar s_{\beta_2}^{b_2}(\eta_2 l) \bar d_{\beta_3}^{b_3}(\eta_3 l) |\bar \Xi^+(\bar P) \rangle \frac{1}{2!} \langle 0| s_{\alpha_1}^{a_1}(\xi_1 n) s_{\alpha_2}^{a_2}(\xi_2 n) d_{\alpha_3}^{a_3}(\xi_3 n) |\Xi^-(P) \rangle \nonumber\\
& \times \frac{-i}{(q - \hat k_2 - \hat p_2)^2} \cdot \frac{-i}{(\hat k_1 + \hat p_1)^2} \times \Bigl[ (-id_s \gamma_5 \sigma^{\mu\nu} q_{\nu}) \frac{i[\gamma\cdot(q - \hat p_2)]}{(q - \hat p_2)^2} (-ig_s\gamma^{\mu_1}) \Bigr]_{\beta_2\alpha_2} (T^a)^{b_2a_2} \nonumber\\
&\times \Bigl[ (-ig_s\gamma^{\mu_2}) \frac{i\gamma\cdot(-\hat k_1 - \hat p_1 - \hat p_3)}{(-\hat k_1 - \hat p_1 - \hat p_3)^2} (-ig_s\gamma_{\mu_1})\Bigr]_{\beta_3\alpha_3} (T^bT^a)^{b_3a_3} \times \Bigl[ -ig_s\gamma_{\mu_2} \Bigr]_{\beta_1\alpha_1} (T^b)^{b_1a_1}.\label{eq:Amplitude}
\end{align}
We have already made a collinear expansion for the hard part in the amplitude. Only the longitudinal components of the parton momenta are kept to get the leading twist contribution, i.e., we take $\hat k_i = (k_i^+,0,\boldsymbol{0}_\perp)$, $\hat p_i = (0,p_i^-,\boldsymbol{0}_\perp)$.
We will denote $k_i^+ = x_i P^+$, and $p_i^- = y_i \bar P^-$, with $x_i$ or $y_i$ the momentum fraction of the hyperon or anti-hyperon carried by the partons.
For each diagram that distinguishes the two strange quarks, the $(1/2!)$ factors in Eq.~(\ref{eq:Amplitude}) compensate the interchange of the two identical strange-quark labels in the three-quark basis.
With the operator-defined LCDAs used here, one such factor accompanies each external hadronic matrix element in the amplitude.
Using the color-singlet structure of the hyperon, the hadronic matrix element can be written as
\begin{align}
\langle 0| s_{\alpha_1}^{a_1}(\xi_1 n) s_{\alpha_2}^{a_2}(\xi_2 n) d_{\alpha_3}^{a_3}(\xi_3 n) |\Xi^-(P) \rangle = \frac{1}{6} \varepsilon^{a_1 a_2 a_3} \varepsilon^{abc} \langle 0| s_{\alpha_1}^{a}(\xi_1 n) s_{\alpha_2}^{b}(\xi_2 n) d_{\alpha_3}^{c}(\xi_3 n) |\Xi^-(P) \rangle,
\end{align}
and similarly for the $\bar \Xi^+$ matrix element.
The overall color factor for the amplitude is then
\begin{align}
C_B = \Bigl(\frac{1}{6}\Bigr)^2 \varepsilon^{a_1 a_2 a_3} \varepsilon^{b_1 b_2 b_3} (T^a)^{b_2a_2} (T^bT^a)^{b_3a_3} (T^b)^{b_1a_1} = \frac{2}{27}.
\label{eq:CF}
\end{align}

Considering all possible gluon connections except those with a three-gluon vertex, there are fourteen diagrams for the photon attaching to each fermion line, and with three fermion lines there are 42 diagrams in total.
Exchanging the two identical $\bar s$ quarks doubles the number of labeled diagrams to 84.
All of these 84 diagrams share the same color factor $C_B$.
Besides, there are two diagrams with a three-gluon vertex for the photon attached to each fixed fermion line.
With three fermion lines and including exchange of the two identical $\bar s$ quarks, there are twelve diagrams with three-gluon vertex in total.
However, all these twelve diagrams vanish, since the three quarks form a color singlet, which gives a vanishing color factor for such three-gluon diagrams.

Using the symmetry properties of the $\Xi^-$ LCDAs in Eqs.~(\ref{eq:symV})-(\ref{eq:symT}), one can find that the contributions from the photon attached on the two $s$ quark lines are the same.
The contribution from diagrams in which the two identical $\bar s$ quarks are exchanged is also the same as that from diagrams without such an exchange.
The calculation is straightforward.
The final result from the quark EDMs contribution is given by the following factorized form
\begin{align}
d_{\Xi^-}(Q) \Big\vert_{\rm EDMs}
&= C_B \frac{(4\pi\alpha_s)^2}{Q^4} \int [d^3x][d^3y] \Biggl\{ d_{s}\Bigl[{\cal V}_{\Xi^-}([x]) + {\cal A}_{\Xi^-}([x]) \Bigr] {\cal T}_{\bar \Xi^+}([y])  {\cal H}_s([x],[y]) \nonumber\\
& + d_d \Bigl[ \Bigl( {\cal V}_{\Xi^-}([x]) {\cal V}_{\bar \Xi^+}([y]) - {\cal A}_{\Xi^-}([x]) {\cal A}_{\bar \Xi^+}([y]) \Bigl) {\cal H}_d([x],[y]) \nonumber\\
&\qquad + \Bigl( {\cal V}_{\Xi^-}([x]) {\cal A}_{\bar \Xi^+}([y]) - {\cal A}_{\Xi^-}([x]) {\cal V}_{\bar \Xi^+}([y]) \Bigr) {\cal C}_d([x],[y]) \Bigr]  \Biggr\}, \label{eq:fromEDMs}
\end{align}
where the hard coefficients are
\begin{align}
{\cal H}_s([x],[y]) &= \frac{-2}{\bar x_2^2 \bar y_2^2}\Bigl( \frac{1}{x_3 y_3} + \frac{1}{x_1 y_1} \Bigr) + \frac{1}{x_1x_3y_1y_3}\Bigl( \frac{1}{\bar x_2 \bar y_1} - \frac{1}{\bar x_3 \bar y_1} + \frac{1}{\bar x_3 \bar y_2} \Bigr), \\
{\cal H}_d([x],[y]) &= \frac{1}{\bar x_3^2 \bar y_3^2 x_1y_1} + \frac{1}{4x_2x_1y_2y_1} \Bigl( \frac{1}{\bar x_1 \bar y_2} -\frac{1}{\bar x_3 \bar y_2} - \frac{1}{\bar x_2 \bar y_3}  \Bigr), \\
{\cal C}_d([x],[y]) &= \frac{1}{4x_2x_1y_2y_1} \Bigl( \frac{1}{\bar x_2 \bar y_3} + \frac{1}{\bar x_1 \bar y_2} - \frac{1}{\bar x_3 \bar y_2} \Bigr).
\end{align}
We have used the notations $\bar x_i \equiv 1-x_i$ and $\bar y_i \equiv 1-y_i$.

\subsection{Quark CEDMs contribution to $d_{\Xi^-}(Q)$}
We proceed to the contributions generated by the quark CEDMs.
A typical diagram for the contribution from the $s$ quark CEDM $\tilde d_s$ is shown in Fig.~\ref{fig:EDM}(b).
In the diagram, one of the quark-gluon vertices is replaced by the CEDM vertex, and the photon vertex is the usual QED vertex.
For the CEDM vertex on each quark line, there are 56 diagrams.
Including the diagrams corresponding to the exchange of the identical $\bar s$ quarks doubles the overall contribution.
Due to the color structure, the contributions from the three-gluon vertex and the two-gluon part of the dipole interactions in Eq.~(\ref{eq:lagrangian}) vanish.
Additionally, because of helicity conservation, neither the gluon CEDM nor the CP-violating Weinberg operator contributes at the leading twist.

The calculation here closely follows that of the quark EDMs case.
The result is given by
\begin{align}
d_{\Xi^-}(Q) \Big\vert_{\rm CEDMs}
&= C_B \frac{e(4\pi \alpha_s)^{3/2}}{Q^4} \int [d^3x] [d^3y] \Biggl\{ \tilde d_{s} \Bigl[{\cal V}_{\Xi^-}([x]) + {\cal A}_{\Xi^-}([x]) \Bigr] {\cal T}_{\bar \Xi^+}([y]) \tilde{\cal H}_s([x],[y]) \nonumber\\
&+ \tilde d_d \Bigl[ {\cal V}_{\Xi^-}([x]) {\cal V}_{\bar \Xi^+}([y]) - {\cal A}_{\Xi^-}([x]) {\cal A}_{\bar \Xi^+}([y]) \Bigr] \tilde {\cal H}_d([x],[y]) \Biggr\}, \label{eq:fromCEDMs}
\end{align}
with the hard coefficient functions
\begin{align}
\tilde{\cal H}_s([x],[y]) &= - Q_s \Bigl[ \frac{2}{\bar x_2 \bar y_2}\Bigl( \frac{1}{x_1 y_1} + \frac{1}{x_3 y_3} \Bigr) + \frac{2}{x_2 y_2}\Bigl( \frac{1}{\bar x_1 \bar y_1} + \frac{1}{x_3 y_3} \Bigr) + \frac{4}{x_3 y_3} \Bigl( \frac{1}{x_1 y_1} + \frac{1}{\bar x_1 \bar y_1} \Bigr) \Bigr] \nonumber\\
&~~~-Q_d \Bigl[ \frac{2}{x_2y_2}\Bigl( \frac{1}{x_1 y_1} + \frac{1}{\bar x_3 \bar y_3} \Bigr) + \frac{4}{x_1 y_1 \bar x_3 \bar y_3} \Bigr], \\
\tilde {\cal H}_d([x],[y]) &= Q_s \Bigl[ \frac{1}{2x_3 y_3}\Bigl( \frac{1}{x_1 y_1} + \frac{1}{\bar x_1 \bar y_1} + \frac{1}{x_2 y_2} + \frac{1}{\bar x_2 \bar y_2} \Bigr) + \frac{1}{x_1 y_1 \bar x_2 \bar y_2} + \frac{1}{ \bar x_1 \bar y_1 x_2 y_2 } \Bigr] \nonumber\\
&+ Q_d \Bigl[ \frac{1}{2\bar x_3 \bar y_3}\Bigl( \frac{1}{x_1 y_1} + \frac{1}{x_2 y_2} \Bigr) + \frac{1}{x_2 x_1 y_2 y_1} \Bigr],
\end{align}
where $Q_s$ and $Q_d$ are the quark electric charges in units of the positron charge $e$.

The total contribution from quark CP-violating dipole interactions to the $\Xi^-$ EDM form factor is then given by the sum of Eqs.~(\ref{eq:fromEDMs}) and (\ref{eq:fromCEDMs}), i.e.,
\begin{align}
d_{\Xi^-}(Q) = d_{\Xi^-}(Q) \Big\vert_{\rm EDMs} + d_{\Xi^-}(Q) \Big\vert_{\rm CEDMs}. \label{eq:XiEDM-result}
\end{align}
We see explicitly that $d_{\Xi^-}(Q)$ scales as $1/Q^{4}$, which agrees with the general power-counting rules for exclusive processes at the leading-twist~\cite{Brodsky:1973kr,Ji:2003fw}.

\subsection{EDM form factor for $\Sigma^+$}
We extend the calculation to $\Sigma^+$ hyperon EDM form factor.
The leading-twist LCDAs for $\Sigma^+$ are defined as~\cite{Chernyak:1987nu}
\begin{align}
& 4\varepsilon^{abc} \int \frac{d\xi_1 d\xi_2 d\xi_3}{(2\pi)^3} e^{iP^+(x_1\xi_1 + x_2\xi_2 + x_3\xi_3)} \langle 0| u_\alpha^a\left(\xi_1 n\right) u_\beta^b\left(\xi_2 n\right) s_\gamma^c\left(\xi_3 n\right)\left|\Sigma^{+}(P)\right\rangle \nonumber \\
&= (\gamma\cdot P C)_{\alpha \beta}\left(\gamma_5 U_{\Sigma^+} \right)_\gamma {\cal V}_{\Sigma^+}\left([x]\right) + \left(\gamma\cdot P \gamma_5 C\right)_{\alpha \beta} \left(U_{\Sigma^+}\right)_\gamma {\cal A}_{\Sigma^+}\left([x]\right) + \left( P_{\nu} i\sigma^{\mu \nu} C\right)_{\alpha \beta}\left(\gamma_\mu \gamma_5 U_{\Sigma^+}\right)_\gamma {\cal T}_{\Sigma^+}\left([x]\right) + \cdots, \label{eq:Sigma-LCDA}
\end{align}
with the valence quark ordering $\Sigma^+(q_1q_2q_3) = \Sigma^+(uus)$. The $\Sigma^+$ LCDAs satisfy
\begin{align}
& {\cal V}_{\Sigma^+}\left(x_1, x_2, x_3\right) = {\cal V}_{\Sigma^+}\left(x_2, x_1, x_3\right), \\
& {\cal A}_{\Sigma^+}\left(x_1, x_2, x_3\right) = -{\cal A}_{\Sigma^+}\left(x_2, x_1, x_3\right), \\
& {\cal T}_{\Sigma^+}\left(x_1, x_2, x_3\right) = {\cal T}_{\Sigma^+}\left(x_2, x_1, x_3\right).
\end{align}
These LCDAs are related to the $\Sigma^+$ state in the helicity basis by~\cite{Bali:2015ykx}
\begin{align}
\left|\Sigma^{+\uparrow}\right\rangle= & \int_0^1 \frac{[d^3x]}{8 \sqrt{6x_1x_2x_3}}\Bigl\{ \Bigl[{\cal V}_{\Sigma^+}([x]) + {\cal A}_{\Sigma^+}([x])\Bigr] \left|u^{\downarrow} u^{\uparrow} s^{\uparrow} \right\rangle 
+ \Bigl[{\cal V}_{\Sigma^+}([x]) - {\cal A}_{\Sigma^+}([x])\Bigr] \left|u^{\uparrow} u^{\downarrow} s^{\uparrow} \right\rangle 
- 2 {\cal T}_{\Sigma^+}([x])\left|u^{\uparrow} u^{\uparrow} s^{\downarrow} \right\rangle\Bigr\}.\label{eq:SigmaWF}
\end{align}
We see that the definitions of LCDAs for $\Xi^-(ssd)$ and $\Sigma^+(uus)$ have exactly the same pattern under the quark flavor replacements of $s$ in $\Xi^-$ by $u$ in $\Sigma^+$, and $d$ in $\Xi^-$ by $s$ in $\Sigma^+$.
The perturbative calculation is also completely analogous to that for $\Xi^-$.
Therefore, the result for $d_{\Sigma^+}(Q)$ can be obtained directly from that of $d_{\Xi^-}(Q)$ in Eq.~(\ref{eq:XiEDM-result}) by substituting the corresponding parameters.
One can get $d_{\Sigma^+}(Q)$ by replacing $(d_s,d_d,\tilde d_s,\tilde d_d,Q_s,Q_d)$ in $d_{\Xi^-}(Q)$ with $(d_u,d_s, \tilde d_u,\tilde d_s,Q_u,Q_s)$, and the simultaneous replacement of the $\Xi^-$ or $\bar \Xi^+$ LCDAs with the $\Sigma^+$ or $\bar \Sigma^-$ LCDAs, respectively.
Here, for brevity, we omit the explicit expressions obtained from the above parameter substitutions.

One can also obtain the EDM form factor results for the other hyperons within the isospin multiplet via similar substitutions.
For example, with the result of $d_{\Xi^-}(Q)$, one can get the EDM form factor for $\Xi^0(ssu)$ with the replacements: $d_d\to d_u$, $\tilde d_d\to\tilde d_u$, and $Q_d\to Q_u$, together with changing the corresponding LCDAs to those of $\Xi^0$ and $\bar\Xi^0$.
Also, the result of $d_{\Sigma^-}(Q)$ can be obtained from $d_{\Sigma^+}(Q)$ with the replacement of flavor parameters from $u$ to $d$ along with the corresponding LCDAs replacement.

\section{Numerical analysis}
\label{sec:numerical}
With the analytical results in the previous section, we now perform a numerical estimation for $d_{\Xi^-}(Q)$ and $d_{\Sigma^+}(Q)$.
Recently, the BESIII Collaboration measured the $\Lambda$ EDM form factor using $J/\psi$ decays~\cite{BESIII:2025vgs}, highlighting this decay channel as a powerful tool for measuring hyperon EDMs.
To confront with possible experimental measurements, we evaluate the EDM form factors at $Q=M_{J/\psi}$.

\subsection{Inputs from hyperon LCDAs}
The numerical analysis requires non-perturbative inputs from the hyperon LCDAs, which have been determined using QCD sum rules~\cite{Chernyak:1987nu,Liu:2008yg,Liu:2014uha} and lattice QCD~\cite{Bali:2015ykx,RQCD:2019hps,Bali:2024oxg,LatticeParton:2024vck,LPC:2025jvd,LPC:2026lcj,LPC:2026mvw}.
Lattice QCD inputs for hyperon LCDAs are available from both the RQCD Collaboration~\cite{Bali:2015ykx,RQCD:2019hps,Bali:2024oxg} and the Lattice Parton Collaboration (LPC)~\cite{LatticeParton:2024vck,LPC:2025jvd,LPC:2026lcj,LPC:2026mvw}.
The RQCD Collaboration provides normalization constants and first conformal moments for all members of the baryon octet with isospin relating the charge states.
We therefore adopt the RQCD inputs throughout to maintain consistency across the hyperons considered.
It is also worth noting that recent studies from the LPC within large-momentum effective theory~\cite{Ji:2013dva,Ji:2014gla,Ji:2020ect} enable access to the LCDAs without relying on moment expansions.
The momentum-fraction dependence of the $\Lambda$ LCDAs has been provided, while corresponding numerical results for $\Xi$ and $\Sigma$ have not yet been reported.
We note that our numerical results are intended as estimates. Although individual coefficients can depend on the LCDA shapes, reasonable variations of these inputs are not expected to change the overall order of magnitude of our predictions.

RQCD uses $\Xi^0$ as the representative of the $\Xi$ isospin doublet. Its parameters apply to $\Xi^-(ssd)$ in the isospin limit.
It should be noted that the explicit phase conventions give them opposite overall signs.
They choose ${\rm DA}^\Xi \equiv {\rm DA}^{\Xi^0} = -{\rm DA}^{\Xi^-}$ where ${\rm DA}^H$ are the LCDAs of hyperon $H$.
Similarly, for $\Sigma$ hyperon, they choose $\Sigma^-$ as the representative, and ${\rm DA}^\Sigma \equiv {\rm DA}^{\Sigma^-} = -{\rm DA}^{\Sigma^+}$.
However, when calculating the EDM form factors, the common sign cancels between the baryon and antibaryon LCDA factors.

Up to the next term of the conformal partial wave expansion, the LCDAs take the following form~\cite{Bali:2015ykx}
\begin{align}
{\cal V}_H([x],\mu)
={}&120x_1x_2x_3\left[
\varphi_{00}^{H}(\mu)
+\frac{7}{2}\bigl(3\varphi_{10}^{H}(\mu)-\varphi_{11}^{H}(\mu)\bigr)
(x_1+x_2-2x_3)\right],\\
{\cal A}_H([x],\mu)
={}&-120x_1x_2x_3\frac{21}{2}
\bigl(\varphi_{10}^{H}(\mu)+\varphi_{11}^{H}(\mu)\bigr)(x_1-x_2),\\
{\cal T}_H([x],\mu)
={}&120x_1x_2x_3\left[
\pi_{00}^{H}(\mu)+7\pi_{11}^{H}(\mu)(x_1+x_2-2x_3)\right],\label{eq:xi-lattice-lcdas}
\end{align}
where $H$ is $\Xi$ or $\Sigma$.
The coefficients $\pi_{nk}^H$ and $\varphi_{nk}^H$ are called as the shape parameters.
The scale evolution of the shape parameters at one-loop is given by~\cite{Bali:2015ykx}
\begin{align}
& \varphi_{nk}^{H}(\mu)
=\varphi_{nk}^{H}(\mu_0)
\left[\frac{\alpha_s(\mu)}{\alpha_s(\mu_0)}\right]^{\gamma_{nk}/\beta_0}, \\
&\pi_{nk}^{H}(\mu)
=\pi_{nk}^{H}(\mu_0)
\left[\frac{\alpha_s(\mu)}{\alpha_s(\mu_0)}\right]^{\gamma_{nk}/\beta_0},\\
& \alpha_s(\mu)=\frac{4\pi}{\beta_0\ln(\mu^2/\Lambda_{\mathrm{QCD}}^2)}, \label{eq:running}
\end{align}
where $\beta_0 = 11-2N_f/3$ with $N_f=3$, $\Lambda_{\mathrm{QCD}}=0.246\,\mathrm{GeV}$, and $\gamma_{00}=2/3$, $\gamma_{10}=26/9$, $\gamma_{11}=10/3$.
We use the central values of the updated determination from RQCD Collaboration for hyperon LCDAs, obtained at the initial scale of $\mu_0=2\,\mathrm{GeV}$.
The shape parameters are (in units of $10^{-3}\,\mathrm{GeV}^2$)~\cite{Bali:2024oxg}
\begin{align}
& \varphi_{00}^{\Xi}= 6.15, \quad \pi_{00}^{\Xi} = 6.34, \quad \varphi_{10}^{\Xi}= 0.361, \quad \varphi_{11}^{\Xi}= -0.001, \quad \pi_{11}^{\Xi}= 0.399, \label{eq:shapeXi}\\
& \varphi_{00}^{\Sigma}= 5.32, \quad \pi_{00}^{\Sigma} = 5.15, \quad \varphi_{10}^{\Sigma}= 0.094, \quad \varphi_{11}^{\Sigma}= 0.200, \quad \pi_{11}^{\Sigma}= -0.080. \label{eq:shapeSigma}
\end{align}
Evolving the shape parameters to $\mu = Q = M_{J/\psi}$, we will get the LCDAs inputs for the numerical calculation of the hyperon EDM form factors.

\subsection{Numerical predictions}
By carrying out the convolution integrals in the factorization formula with the inputs of hyperon LCDAs, we obtain
\begin{align}
d_{\Xi^-}(M_{J/\psi}) &= 
-8.67\times10^{-4} d_s
-1.19\times10^{-6} d_d +3.46\times10^{-4} e\tilde d_s
-7.16\times10^{-5} e\tilde d_d,\label{eq:XiEDM-num} \\
d_{\Sigma^+}(M_{J/\psi})&=
4.24\times10^{-5}\,d_s - 3.20\times10^{-4}\,d_u + 4.42\times10^{-5}\,e\tilde d_s -2.50\times10^{-4}\,e\tilde d_u. \label{eq:SigmaEDM-num}
\end{align}
The numerical results exhibit a hierarchy among the flavor contributions. 
At $Q=M_{J/\psi}$, the coefficients multiplying the strange-quark EDM and CEDM in $d_{\Xi^-}$ are significantly larger than the corresponding down-quark coefficients in magnitude. Similarly, in $d_{\Sigma^+}$, the up-quark EDM and CEDM coefficients exceed their strange-quark counterparts significantly. 
Comparing the two hyperons, the strange-quark EDM and CEDM coefficients in $d_{\Xi^-}$ are about one order of magnitude larger than those in $d_{\Sigma^+}$. 
These patterns reflect the valence compositions of $\Xi^-(ssd)$ and $\Sigma^+(uus)$, as well as their spin-flavor structure and LCDA shapes.
From this numerical result, we see that $\Xi^-$ provides greater sensitivity to strange-quark dipole interactions than $\Sigma^+$.

With the EDM bounds for the neutron and the $\Lambda$ hyperon, the constraints for the strange-quark EDM and CEDM have been derived in~\cite{Chen:2025rab}.
The constraints are obtained in two scenarios.
Scenario-I assumes a hierarchy with $d_s \gg d_u, d_d$; Scenario-II adopts a ${\rm SU}(3)$ flavor-symmetric model with $d_s=d_u=d_d$.
In both scenarios, $\tilde d_u$ and $\tilde d_d$ are neglected, so that $\tilde d_s$ is the only nonzero quark CEDM.
We note that there is a sign-convention problem in~\cite{Chen:2025rab}.
The quark EDM contribution to $d_\Lambda$ requires an additional overall minus sign, while the quark CEDM contribution remains unchanged.
The stringent neutron EDM constraint makes this correction
numerically negligible in both scenarios, leaving the
allowed regions in Fig.~2 and the obtained $\tilde d_s$
bounds of that reference unchanged at the displayed precision.
Here, we retain these two scenarios and obtain
\begin{align}
& d_{\Xi^-}^{\rm I}(M_{J/\psi})
\simeq -8.67\times10^{-4}d_s
+3.46\times10^{-4}e\tilde d_s, \label{eq:XiEDM-s1}\\
& d_{\Xi^-}^{\rm II}(M_{J/\psi})
\simeq -8.68\times10^{-4}d_s
+3.46\times10^{-4}e\tilde d_s, \label{eq:XiEDM-s2}
\end{align}
for $\Xi^-$, and 
\begin{align}
& d_{\Sigma^+}^{\rm I}(M_{J/\psi})
\simeq4.24\times10^{-5}d_s
+4.42\times10^{-5}e\tilde d_s, \label{eq:SigmaEDM-s1}\\
& d_{\Sigma^+}^{\rm II}(M_{J/\psi})
\simeq -2.77\times10^{-4}d_s
+4.42\times10^{-5}e\tilde d_s.\label{eq:SigmaEDM-s2}
\end{align}
for $\Sigma^+$.
The superscripts ``${\rm I/II}$" denote for scenario-I/II.

The bound for $s$ quark EDM is obtained using the neutron EDM bound~\cite{Abel:2020pzs}.
It is, in two scenarios~\cite{Chen:2025rab}
\begin{align}
& |d_s^{\rm I}|<6.67\times10^{-24}\,e\,{\rm cm},\qquad
|d_s^{\rm II}|<3.09\times10^{-26}\,e\,{\rm cm}. \label{eq:ds-bound}
\end{align}
The bound for $s$ quark CEDM is derived with the additional bound on the $\Lambda$ EDM.
To the current accuracy, either scenario gives,
\begin{align}
-1.38\times10^{-14}\,{\rm cm} < \tilde d_s < 5.31\times10^{-15}\,{\rm cm}. \label{eq:tdsbound}
\end{align}
Inserting the bounds in Eqs.~(\ref{eq:ds-bound}) and (\ref{eq:tdsbound}) into Eqs.~(\ref{eq:XiEDM-s1})-(\ref{eq:SigmaEDM-s2}), we get, for both scenarios to the displayed accuracy, the ranges of the hyperon EDM form factor
\begin{align}
& -4.8\times10^{-18} e\,{\rm cm}
< d_{\Xi^-}(M_{J/\psi}) 
< 1.8\times10^{-18} e\,{\rm cm}, \label{eq:XiEDM-bound} \\
& -6.1\times10^{-19}\,e\,{\rm cm}
< d_{\Sigma^+}(M_{J/\psi})
< 2.4\times10^{-19}\,e\,{\rm cm}. \label{eq:SigmaEDM-bound}
\end{align}
The ranges of $d_{\Xi^-}$ and $d_{\Sigma^+}$ are both primarily controlled by $\tilde d_s$ bound, which indicates that the future measurement of these hyperon EDMs will play a decisive role in constraining the range of $s$ quark CEDM.
These measurements are promising to be carried out on BESIII. Further improvements are also expected at the planned Super Tau-Charm Factory~\cite{Achasov:2023gey}.

Before giving our summary, we note that higher-order $\alpha_s$ corrections, $1/Q$-suppressed higher-twist and hadron-mass corrections are neglected in this analysis.
Recent next-to-leading-order calculations of nucleon form factors (see e.g.,~\cite{Chen:2024fhj,Huang:2024ugd}) demonstrate that radiative corrections to hard-exclusive amplitudes can be sizable.
Although these corrections could be significant at the moderate scale of $Q=M_{J/\psi}$, we neglect them here and leave their investigation to future works.
Furthermore, the hyperon EDM form factor is, in general, complex in the timelike region. Our tree-level calculation at the leading-twist determines its real contribution, and no absorptive phase is calculated.

\section{Summary}
\label{sec:summary}
In this paper, we have calculated the EDM form factor of $\Xi^-$ and $\Sigma^+$ within the perturbative QCD factorization framework for hard exclusive processes. 
We have included both quark EDMs and CEDMs contributions and obtained the form factors at the leading-order and leading-twist in terms of the convolution of the hard coefficients and hyperon LCDAs.
Using the lattice QCD inputs for the hyperon LCDAs, we obtain the numerical dependence of hyperon EDMs on quark dipole couplings. 
With the constraints of strange quark dipole couplings derived from the neutron and $\Lambda$ hyperon EDMs, we give predictions for the bounds of the $\Xi^-$ and $\Sigma^+$ EDM form factors.
The bounds predicted are expected to be accessible via current BESIII experiment.
With the new physics programs at BESIII~\cite{Zhang:2026pyv} and the planned Super Tau-Charm Factory~\cite{Cheng:2025kpp}, further constraints on the CP-violating dipole interactions at the quark level beyond the SM are expected.
There are also great opportunities for studying quark dipole interactions at the future electron-ion colliders~\cite{Boughezal:2023ooo,Wen:2024cfu,Huang:2025ljp} and lepton colliders~\cite{Wen:2024nff,Cao:2025qua,Yu:2026cdd}.

\section*{Acknowledgments}
K.-B. C. would like to thank Shou-Shan Bao for useful discussions.
This work is supported by National Natural Science Foundation of China (Grant No. 12005122), Shandong Province Natural Science Foundation (Grant No. ZR2020QA082), and Youth Innovation Team Program of Higher Education Institutions in Shandong Province (Grant No. 2023KJ126).
J.-P. M. is supported by National Key R\&D Program of China Grants No. 2024YFE0109800 and No. 2024YFE0109801.
X.-B. T. is supported by the Research Council of Finland, the Centre of Excellence in Quark Matter and Projects No. 338263 and No. 359902, and by the European Research Council (ERC, Grant Agreements No. ERC-2023-101123801 GlueSatLight and No. ERC-2018-ADG-835105 YoctoLHC).

\end{document}